\documentclass[twocolumn,floatfix,superscriptaddress,amsmath,showpacs,showkeys,aps,pre]{revtex4-2}

\usepackage{environ,censor,calc}
\NewEnviron{mysout}
 {\soutrefunexp{\BODY}}
\newlength\nextcharwidth
\makeatletter
\renewcommand\@cenword[1]{%
  \setlength{\nextcharwidth}{\widthof{#1}}%
  \censorrule{\nextcharwidth}%
  \kern -\nextcharwidth%
  #1}
\makeatother

\usepackage{epsfig}
\usepackage{bm}
\usepackage[dvipsnames]{xcolor}
\usepackage{amsmath}
\usepackage{amssymb}
\usepackage{graphicx}
\usepackage{physics}
\usepackage{xr}
\usepackage{soul}
\usepackage{longtable}
\usepackage{verbatim}
\usepackage[OT1]{fontenc}
\usepackage[normalem]{ulem}
\usepackage{blkarray}
\usepackage{hyperref}
\usepackage{multirow}

\def\visualizar{2}

\newcommand\new[1]{\textcolor{blue}{#1}}

\if\visualizar1
\renewcommand\old[1]{}
\fi
  \if\visualizar0
 \renewcommand\new[1]{}
\fi
\if\visualizar3
 \renewcommand\new[1]{#1}
 \renewcommand\old[1]{}
\fi

\begin{document}

\title{Finite-time scaling on low-dimensional map bifurcations}

 \author{Daniel A. Martin}
 \affiliation{Instituto de Ciencias F\'isicas (ICIFI-CONICET), Center for Complex Systems and Brain Sciences (CEMSC3), Escuela de Ciencia y Tecnolog\'ia, Universidad Nacional de Gral. San Mart\'in, Campus Miguelete, San Mart\'in, Buenos Aires, Argentina}
\affiliation{Consejo Nacional de Investigaciones Cient\'{\i}fcas y Técnicas (CONICET), Buenos Aires, Argentina}
\author{Qian-Yuan Tang}
\affiliation{Department of Physics, Faculty of Science, Hong Kong Baptist University, Hong Kong SAR, China }

\author{Dante R. Chialvo}
\affiliation{Instituto de Ciencias F\'isicas (ICIFI-CONICET), Center for Complex Systems and Brain Sciences (CEMSC3), Escuela de Ciencia y Tecnolog\'ia, Universidad Nacional de Gral. San Mart\'in, Campus Miguelete, San Mart\'in, Buenos Aires, Argentina}
\affiliation{Consejo Nacional de Investigaciones Cient\'{\i}fcas y Técnicas (CONICET), Buenos Aires, Argentina}
\affiliation{Department of Physics, Faculty of Science, Hong Kong Baptist University, Hong Kong SAR, China }

\date{\today}

\begin{abstract}
Recent work has introduced the concept of finite-time scaling to characterize bifurcation diagrams at finite times in deterministic discrete dynamical systems, drawing an analogy with finite-size scaling used to study critical behavior in finite systems. In this work, we extend the finite-time scaling approach in several key directions. 
First, we present numerical results for 1D maps exhibiting period-doubling bifurcations and discontinuous transitions, analyzing selected paradigmatic examples. We then define two observables, the finite-time  susceptibility and the finite-time Lyapunov exponent, that also display consistent scaling near bifurcation points. The method is further generalized  to special cases of 2D maps including the 2D Chialvo map, capturing its bifurcation between a fixed point and a periodic orbit, while accounting for discontinuities and asymmetric periodic orbits. 
These results underscore fundamental connections between temporal and spatial observables in complex systems, suggesting new avenues for studying complex dynamical behavior.
\end{abstract}

 \maketitle

\section{Introduction}

Dynamical systems, including physical, ecological and neural models, can often be described by low-dimensional maps \cite{may,ott,kaneko2,glass, kaneko,review1,review2}. These maps are analytically more tractable and computationally faster to simulate than their high-dimensional, spatially extended counterparts. A critical feature of such maps is the occurrence of bifurcations—parameter values where the system's evolution undergoes qualitative changes.

While the asymptotic (infinite-time) behavior near bifurcations has been extensively studied, much less is known about how systems approach these regimes over finite time—yet this transient behavior is essential in many real-world contexts, from biological rhythms to control processes.
Recent work by Corral and collaborators~\cite{corral1,corral2,falco} introduced the concept of \emph{finite-time scaling} in one-dimensional maps. Drawing an analogy with finite-size scaling in statistical physics, they demonstrated that as the number of iterations increases, observables near bifurcation points exhibit scaling collapses governed by a joint dependence on iteration time and distance from criticality. This framework was first developed for the Galton–Watson process~\cite{WG1,WG2} and later extended to several classes of 1D maps and ordinary differential equations exhibiting continuous bifurcations~\cite{corral2}. This approach highlights how finite-time dynamics near critical points can carry universal signatures, much like spatially extended systems approaching thermodynamic phase transitions.

In this article, we extend finite-time scaling analysis in two main directions. First, we revisit 1D maps exhibiting period-doubling bifurcations. We also introduce two finite-time observables—the susceptibility and the Lyapunov exponent—and demonstrate that both exhibit characteristic scaling behavior near bifurcations. We later consider discontinuous (hysteretic) bifurcations, including systems where spinodals replace critical points. We then generalize the approach to two-dimensional maps, considering a simple case and also the Chialvo neuron map~\cite{Chialvomap}, which displays diverse bifurcation scenarios.
  
Across models, we interpret finite-time scaling as a dynamical signature of local instability. Low-dimensional observables—such as map iterates—act as effective order parameters that encode the collective dynamics of an underlying high-dimensional system. Finite-time scaling thus serves as a dynamical analogue of critical behavior near bifurcations.

The article is organized as follows: Section~\ref{model} reviews the basic formulation of finite-time scaling, outlines its extension to various bifurcation types and define additional observables relevant for scaling behavior. Section~\ref{results} presents results for both 1D and 2D maps. For completeness, technical derivations are included in Appendix. We conclude in Section~\ref{conclusion} with a discussion of the implications for dynamical systems and complex behavior.

\section{Methods and models}
\label{model}
\subsection{Finite-time scaling approach}

In their seminal article, Corral et al.~\cite{corral1} introduced a finite-time scaling framework for one-dimensional maps, demonstrating that their behavior near bifurcations is analogous to finite-size scaling in extended systems approaching continuous phase transitions. This includes critical slowing down at the bifurcation point, critical exponents, curve collapses, and related phenomena---with the system size $L$ effectively replaced by the number of iterations $l$, and the control parameter $(m - 1)$ analogous to a reduced temperature $(T - T_c)$ in phase transitions.

Consider a one-dimensional map $x_{l+1}=f_{\mu}(x_l)$, having a  fixed point at $x=p$. The function $f$ and the fixed point $p$ depend on a bifurcation parameter $\mu$, and $f$ can be expanded as
\begin{equation}
    f_{\mu}(x)
    =p+m_{\mu} (x - p) - P (x - p)^k+O[(x - p)^{k+1}].\label{general}
\end{equation}
Here, $m_{\mu} = f'_{\mu}(p)$ quantifies linear stability, $P$ controls the leading nonlinearity, and $k \geq 2$ determines the dominant curvature near the fixed point. In what follows, we treat $\mu$ as the fundamental control parameter of the bifurcation, while $m = f'_\mu(p)$ denotes the local linear stability (slope) evaluated at the fixed point $p$. For convenience, we express the scaling variable and other relations in terms of $m$, although it is implicitly a function of $\mu$.

In Ref.~\cite{corral1}, the authors applied this framework to three one-dimensional cases (specifically,  the first bifurcation point of the logistic map, the transcritical map, and the saddle-node) for $k=2$, about $m_{\mu^*} = 1$, where $\mu=\mu^*$ at the bifurcation point \cite{corral1}, the corresponding fixed point at $\mu^*$ is $p^*$, satisfying $f_{\mu^*}(p^*) = p^*$, and $P^*=P_{\mu^*}$.

The same analysis was later extended \cite{corral2} to other one dimensional systems for $k\geq 2$, still about $f'(p^*) = m_{\mu^*} = 1$. About the transition, these maps can be written as:
\begin{equation}
    f_{\mu}(x)=(1+\mu)x-P^* x^k.
    \label{mapkold}
\end{equation}
where $P^*$ controls the strength of the nonlinear term, and $k \geq 2$ determines its leading order. Near the bifurcation point $\mu = 0$, this map captures the interplay between linear growth and nonlinear saturation.

For systems close to the bifurcation point (i.e., $m = 1$, or $\mu = 0$), one can also derive the finite-time scaling after $l \gg 1$ steps:
\begin{equation}
(x_l-p)^{(k-1)} = \frac{ G((k-1)z)}{(k-1) l P^*},
\label{Eqk}
\end{equation}
where $z = l(m - 1)$ is the scaling variable that governs the crossover between critical and off-critical behavior, analogous to $(T - T_c)L^{1/\nu}$ in statistical physics. The scaling function $G(z) = \frac{ze^z}{e^z - 1}$ captures this transition. Further details on the assumptions and limitations can be found in the respective articles \cite{corral1,corral2}.

Eq.~\ref{Eqk} enables a finite-time scaling collapse by expressing $l(p - x_l)$ as a function of the scaling variable $z$, and also characterizes the convergence toward the stable fixed point: at criticality, the deviation decays as a power law, $(x_l - p) \sim l^{-1/(k-1)}$, while away from it, the decay is exponential, $(x_l - p) \sim \exp(-l/\tau)$, with time scale $\tau = 1/|m - 1|$ reflecting critical slowing down. The authors in \cite{corral1,corral2} also find that similar collapses hold for the distance to the bifurcation point, $p*$. Although originally developed for discrete maps, the framework naturally extends to continuous-time systems. By interpreting differential equations as maps with infinitesimal time steps, one can derive analogous scaling relations for the approach to instability.

In all studied cases, the bifurcation is continuous, where one fixed point becomes unstable and another stable at the same critical parameter $\mu^*$. These transitions are non-oscillatory, with the Jacobian eigenvalue at the fixed point remaining real and satisfying \( f'(q^*) = 1 \).

\subsection{Finite-time scaling for period-doubling bifurcations}

We extend the previous results to several cases of interest. First, study period-doubling bifurcations where $f'(p^*) = m_{\mu^*} = -1$.     
 We first consider the maps of the form

\begin{equation}
    f_{\mu}(x) = -(1 + \mu)x + Q x^\kappa, \label{mapk}
\end{equation}
where $\kappa \geq 2$ denotes the exponent of the original nonlinear term. Importantly, the exponent $\kappa$ does not always coincide with the scaling exponent $k$ that governs finite-time behavior. As shown in the Appendix, the effective exponent $k$ depends on the symmetry of the map and may arise from the second iterate $f(f(x))$ when $\kappa$ is even. For period-doubling bifurcations the fixed point becomes unstable at $m = -1$, and the correct distance to criticality is $m + 1$. Accordingly, we define the scaling variable as $z = l(m + 1)$ in these cases to preserve collapse consistency.

The maps in Eq.~\ref{mapk} have a stable solution $x = 0$ for $\mu < 0$~\cite{footnote}. We focus on the cases where $Q > 0$ and/or $\kappa$ is even, for which a period-2 solution emerges when $\mu > 0$. For $\kappa$ odd and $Q < 0$, the solutions diverge, and this case is therefore excluded from the present analysis.

For $\kappa$ odd and $\mu > 0$, the period-2 solutions are $\pm \sqrt[k - 1]{\mu / Q}$.  As shown in the Appendix, the distance to the origin follows Eq.~\ref{Eqk}, provided we take $k = \kappa$ and $z = l(m + 1)$ (instead of $l(m - 1)$). We also find that the distance to the closest period-2 solution obeys a similar scaling law.

For $\kappa$ even and $\mu > 0$, the period-2 solutions are no longer symmetric about zero. Nonetheless, Eq.~\ref{Eqk} still applies, as long as we take $k = 2\kappa - 1$, $z = (m + 1)l$, and $P^* = Q^2(\kappa - 1)$. The derivation of these relations is provided in the Appendix. Numerically, we find that the resulting collapses remain valid, provided we compute the distance to the midpoint of the period-2 solutions rather than to the fixed point $p^*$. That is, we replace $p \to p_m = \frac{p_1 + p_2}{2}$, where $p_1$ and $p_2$ are the two period-2 points satisfying $f(p_1) = p_2$ and $f(p_2) = p_1$.

\subsection{Additional observables and scaling behavior}

In statistical physics, critical phenomena are characterized not only by diverging correlations but also by universal response functions such as susceptibilities. These functions capture how strongly systems react to perturbations and are tightly linked to fluctuations via fluctuation–dissipation theorems. Inspired by this analogy, we explore whether dynamical systems near bifurcations exhibit similar finite-time scaling in their response behavior. Specifically, we consider two observables: the \emph{finite-time susceptibility}, which quantifies sensitivity to parameter changes, and the \emph{finite-time Lyapunov exponent} \cite{ott}, which measures local trajectory instability. We analyze their scaling in the logistic map and the reduced Chialvo neuron model.

\vspace{0.5em}
\noindent\textbf{Definitions.}
We define the \emph{finite-time susceptibility} as
\begin{equation}
\chi_l(\mu) = \sum_{i=1}^{l} \left( \frac{\partial x_i(\mu)}{\partial \mu} \right)^2,
\label{eq:chi_def}
\end{equation}
which captures the total squared response of a length-$l$ trajectory to small perturbations in the control parameter $\mu$. Physically, this quantity is analogous to integrated susceptibility in thermodynamic systems and, under Gaussian assumptions, corresponds to the Fisher information.
For maps described by Eq. \ref{mapkold}, deriving Eq. \ref{Eqk} with respect to $\mu$, it is straightforward to find $\frac{\partial x_l}{\partial \mu} \sim l^\frac{k-2}{k-1}$, then, we expect
\begin{equation}
\chi_l(\mu) \sim l^{(3k - 5)/(k - 1)}. \label{EqChi}
\end{equation}
We find that the same equation holds for simple period doubling bifurcations described by Eq. \ref{mapk} provided we replace $k=\kappa$ (if $\kappa$ is odd) or $k=2 \kappa -1$ if $\kappa$ is even.  

The \emph{finite-time Lyapunov exponent} is defined by
\begin{equation}
\lambda_l = \frac{1}{l} \sum_{i=0}^{l-1} \log \left| f'(x_i) \right|,
\label{eq:lyap_def}
\end{equation}
and characterizes the average local stretching or contraction rate of the system. This reflects the local stability of trajectories and provides a dynamical analogue of relaxation time near bifurcation.

Deriving Eq. \ref{mapkold} with respect to $x$,  recalling that at the bifurcation point $x_l \sim l^{-1/(k-1)}$, and considering that $\log(1+\epsilon)\simeq \epsilon$ for small enough $\epsilon$, we find that: 
\begin{equation}
\lambda_l \sim \frac{1}{l} \sum_{i=0}^{l-1}  l^{-1} \simeq \log(l)/l,
\label{eq:lyap_aprox}
\end{equation}
 
for all values of $k$ (or $\kappa$). We have verified that Eq. \ref{eq:lyap_aprox} holds for simple maps with period doubling bifurcations (given by Eq. \ref{mapk}) and without period doubling bifurcations (Eq. \ref{mapkold}) irrespective of the value of $\kappa$ or $k$. The scaling of Lyapunov exponent includes a logarithmic correction and does not yield a pure power law. This suggests slower divergence in trajectory instability compared to the susceptibility.

In summary, the finite-time susceptibility $\chi_l$ and Lyapunov exponent $\lambda_l$ provide distinct yet complementary perspectives on dynamical criticality. $\chi_l$ reflects sensitivity to external perturbations and follows clear power-law scaling, while $\lambda_l$ captures internal trajectory divergence and may includes logarithmic corrections. Together, they reinforce the analogy between bifurcations and critical phenomena, showing that scaling applies not only to state evolution but also to dynamical responses.

\section{Results}\label{results}
 
\subsection{1D period-doubling bifurcation}
\subsubsection{Simple Map}

We have verified the above mentioned relations among $k$ and $\kappa$, {$z$} and $m$, {$p$ and $p_{1,2}$}  for $Q=1$, $\kappa=3,...,8$.
The finite-time scaling for the map in Eq.(\ref{mapk}), for $\kappa=3$ and $\kappa=8$,  are shown in the Appendix.

Both for $\kappa$ even and odd, we evaluate $m$ about $x=0$. That is, we consider $m=-1-\mu$, so that $z=\mu l$ {and the bifurcation is about $\mu=0$. A very good collapse for different values of $l$ if found where $x l^{1/\kappa-1}$ (for $\kappa$ odd) or $(x-p_m)  l^{1/2\kappa-2}$ (for $\kappa$ even) is plotted as a function of $z$. }
This collapse confirms the finite-time scaling hypothesis for period-doubling bifurcations, with deviations following the scaling law governed by \( z = l(m - 1) \). The consistency across values of \(l\) reflects the expected slowing down near criticality, in agreement with Eq.~\ref{Eqk}.

\begin{figure}[ht!]
    \centering
    \includegraphics[width=0.9\linewidth]{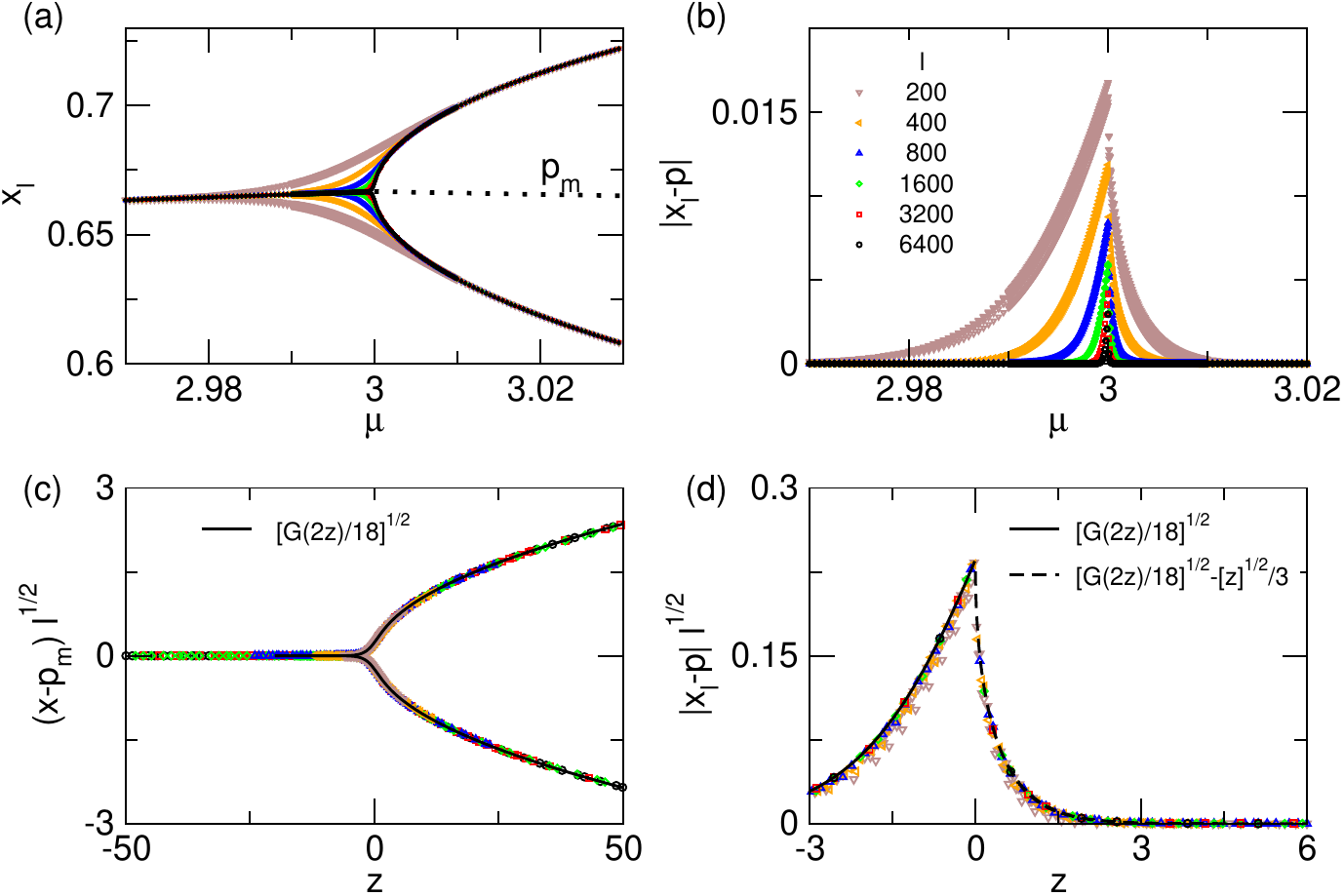}
    \caption{Results for the period-doubling bifurcation of the logistic map (at $\mu=3$). Panel a: solutions $x_l$ for different values of $l$, as a function of $\mu$. A dashed line shows the average of the solutions, $p_m=\frac{p_1+p_2}{2}=\frac{\mu+1}{2\mu}$. Panel b: Distance to the closest fixed point $p_{1/2}=\frac{1}{2\mu}  \left(\mu+1\pm \sqrt{(\mu-3) (\mu+1)}\right)$ Panel c: collapse of the distance to the midpoint, $p_m$, as a function of $z\doteq (\mu-3) l$. Panel d: Collapse of the distance to the closest function. In panels (c) and (d), analytical expressions are plotted in black continuous line.}
    \label{fig:log3}
\end{figure}

\subsubsection{Logistic Map for $\mu=3$}

Probably, the most frequently studied route to chaos is period-doubling. One of the most common maps showing those transitions is the Logistic map, which exhibits a sequence of bifurcations as the parameter $\mu$ increases, with each transition characterized by the emergence of a new stable fixed point and a corresponding change in the convergence rate. The map is given by:
\begin{equation}
    f_{\mu}(x)=\mu x (1-x).
\end{equation}

{For $\mu<1$ the stable solution is $x_1=0$}. The solution $x=0$ loses its stability towards $x_2=\frac{\mu-1}{\mu}$ at $\mu\geq 1$. The bifurcation at $\mu=1$ has already been considered by \emph{Corral et al} \cite{corral1}.

The stationary solution $x_2$ is stable in the range $1<\mu<3$ (initial conditions decay exponentially towards $x_2$ for $1<\mu<2$, where $0<f'(\mu)=2-\mu<1$, and decay oscillating for $2<\mu<3$, where $-1<f'(\mu)<0$). At $\mu=3$,  a period doubling bifurcation ($f'(\mu)=-1$)  takes place. There, the stationary solution $x_2$ becomes unstable. 

We  consider the collapse about the bifurcation at $\mu=3$. Here, $\kappa=2$, consequently $k=3$, $p^*=2/3$, $Q=3$ and $P^*=9$.   The solutions for $\mu>3$ are known analytically (see e.g. \cite[Chapter 10]{Strogatz}): $p_{1/2}=\frac{1}{2\mu}  \left(\mu+1\pm \sqrt{(\mu-3) (\mu+1)}\right)${, so $p_m=\frac{\mu+1}{2\mu}$. The map for different number of iterations together with the collapse to the distance to $p_m$ is shown in Fig. \ref{fig:log3} (a) and (c) respectively. The distance to the closest solution, together with its collapse, are shown in panels (b) and (d).}

\begin{figure}[ht!]
    \centering
    \includegraphics[width=0.975\linewidth]{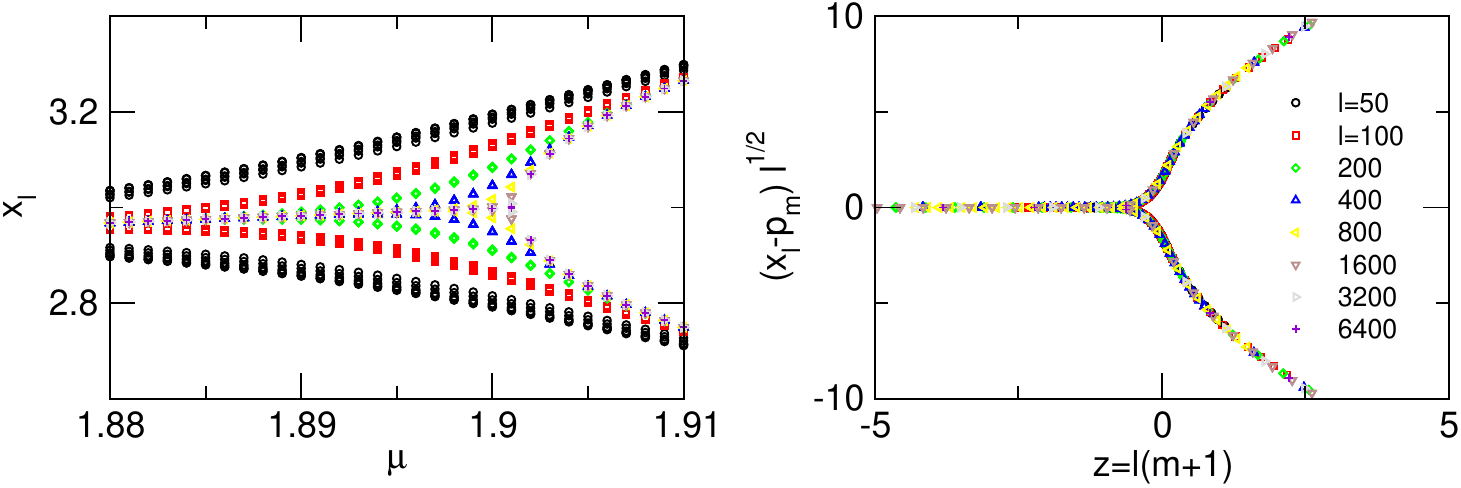}
    \caption{Results for 1D Chialvo Map near the period-doubling transition at $\mu^*=3-\log(3)$, for which $p^*=3$. The left panel corresponds to the original data, while the right panel corresponds to the collapsed one.}
    \label{fig:Ch1d}
\end{figure}

\subsubsection{Reduced 1D Chialvo neuron map}

The Chialvo map \cite{Chialvomap} is a 2D map describing the generic dynamics of excitable systems.  While the full model is given in the next section, we focus here on a 1D approximation valid in the absence of voltage-dependent inactivation and external bias, reducing the system to a single-parameter map:
\begin{equation}
    x_{l+1}=x_l^2e^{\mu-x_l},
    \label{Chialvo1D}
\end{equation}
where parameter $\mu$ controls the system’s excitability. This map has a stable solution  $x=0$ for $\mu<1$, for $\mu>1$, it has  another fixed stable  solution, and  it is straightforward \cite{footnote2}
 to show that at $\mu^*=3-\log(3)\simeq 1.901387711$, for which $p^*=3$, there is a  period doubling bifurcation. About this point, it is similar to the simple map perturbation with $k=2$, and the curves collapse as expected, see Fig. \ref{fig:Ch1d}.

\subsubsection{Additional Observables}
\begin{figure}[ht!]
    \centering
    \includegraphics[width=0.9\linewidth]{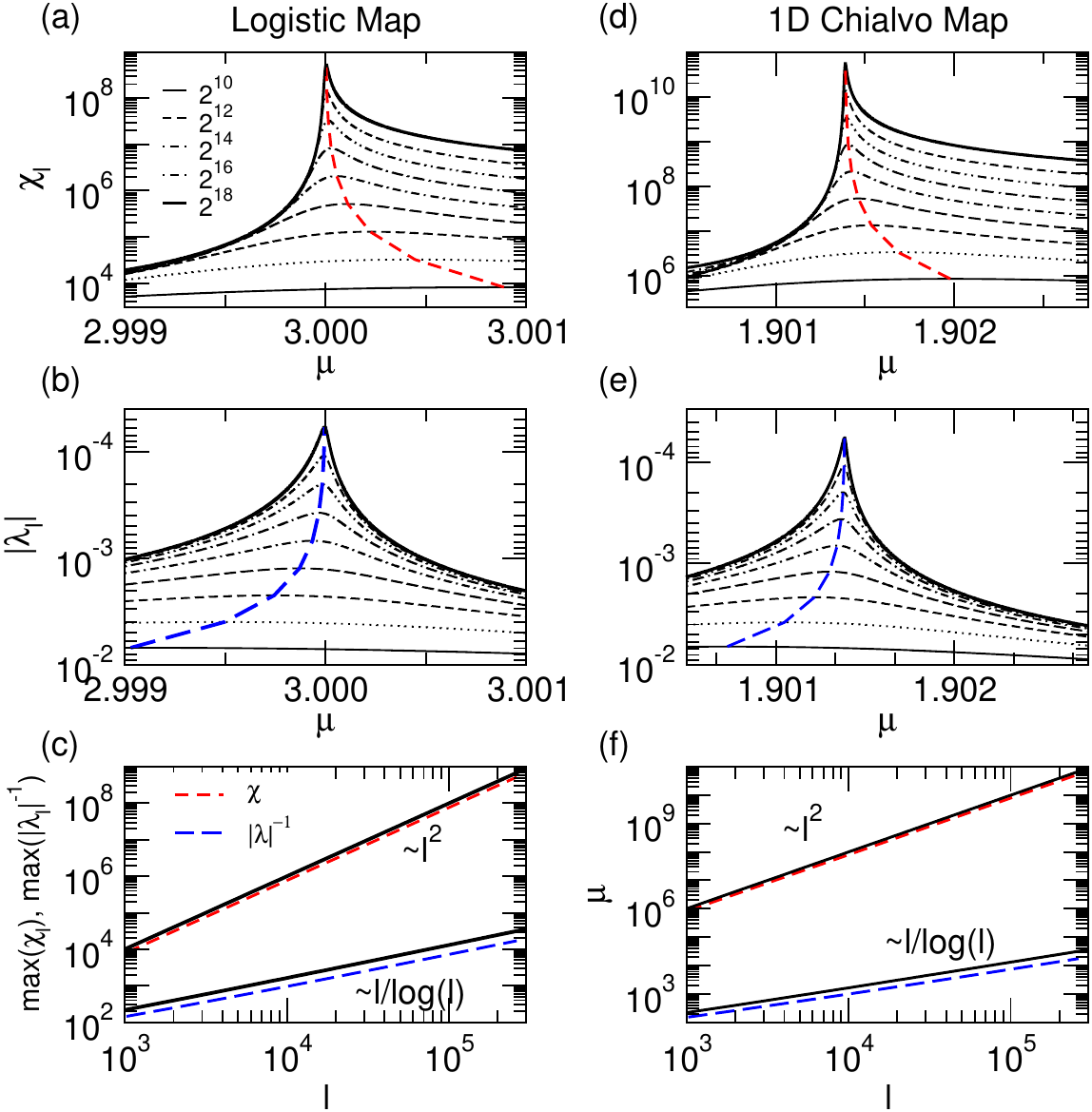}
    \caption{Finite-time susceptibility and Lyapunov exponent for Logistic and 1D Chialvo Maps. Panel a: Finite-time susceptibility $\chi_l$ as a function of $\mu$ for the Logistic map about $\mu=3$, for different values of the number of iterations, $l=2^{10},2^{11},.., 2^{18}$. A red dashed curve joins the maximum $\chi_l$ values and the corresponding $\mu^*$ values. Panel b: Absolute value of the  Finite-time Lyapunov exponent for the Logistic map, for the same values of $l$ as in panel a. A blue dashed line joins the maximum values of $|\lambda|$. Notice that the $y$-scale is inverted. Panel c: Maximum values of $\chi_l$ and $|\lambda_l|^{-1}$ as a function of $l$. Black continuous lines show the expected behavior: $\max(\chi_l) \propto l^{\gamma/\nu}$ with $\gamma/\nu=2$ and 
    $max(\lambda_l)=\min(|\lambda_l|) \propto \log(l)/l$. Panels d-f: same results for the Chialvo 1D map. Initial conditions: $x_0=0.03$ for the Logistic map and $x_0=2.5$ for 1D Chialvo map. }
    \label{fig:Exp1}
\end{figure}

In Fig.~\ref{fig:Exp1}(a), we show the finite-time susceptibility $\chi_l$ for the period-doubling bifurcation of the logistic map, evaluated at various values of $l$. The susceptibility increases with $l$, and for each $l$ it exhibits a maximum at a parameter value $\mu_l^*$. As $l$ increases, we observe that $\mu_l^* \to \mu^* = 3$.

Panel (b) shows the absolute value of the finite-time Lyapunov exponent $\lambda_l$, computed for the same values of $\mu$ and $l$ as in panel (a). Similar to the susceptibility, $\lambda_l$ is negative and increases (i.e., becomes less negative) with $l$, approaching zero at $\mu^*$ as $l \to \infty$.

In panel (c), we plot the maximum values of $\chi_l$ and $\lambda_l$ as functions of $l$. For the logistic map at $\mu = 3$, which undergoes a period-doubling bifurcation with $\kappa = 2$, we expect $k = 2\kappa - 1 = 3$, yielding $\chi_l \sim l^{(3k - 5)/(k - 1)} = l^2$ and $\lambda_l \sim \log(l)/l$. The black solid lines indicate these predicted scalings and show excellent agreement with the observed data.

Panels (d)–(f) show the same analysis for the 1D Chialvo map near its bifurcation at $\mu^* = 3 - \log(3)$. Although the Chialvo map includes an exponential term and differs structurally from Eq.~\ref{mapk}, it can be approximated by a second-degree polynomial near the bifurcation. Thus, we again have $k = 3$, $\kappa = 2$, and the same scaling exponents apply to both $\chi_l$ and $\lambda_l$.

The results in Fig.~\ref{fig:Exp1} exhibit a striking parallel with finite-size scaling behavior in statistical thermodynamics. For both models (using terminology from that context), we find that $\max(\chi) \propto l^{\gamma/\nu}$ with $\gamma/\nu = \frac{3k - 5}{k - 1}$. In the Appendix, we show that $|\mu_l^* - \mu^*| \propto l^{-1/\nu}$ with $\nu = 1$. Furthermore, all curves collapse onto a single one when plotted as $\chi_l l^{-\gamma/\nu}$ versus $(\mu - \mu^*) l^{1/\nu}$, see Fig. \ref{Colapse}.

\subsection{Discontinuous transitions and discontinuous 1D maps bifurcations}

Up to here, the above results exhibits the similarities between  finite time scaling in maps and continuous phase transitions in extended systems. However, dynamical systems may present additional type of transitions, being discontinuous transitions the most common among them.  Discontinuous transitions, such as melting or vaporization in water, are characterized by a discontinuous jump in the order parameter (such as density), the existence of hysteresis and a range of the control parameter space where two phases may take place (one stable and the other metastable, switching stability at the transition point). The (meta)stability limit of those regions are two  spinodals (one for each phase), after which the metastable solutions cease to exist. Spinodals  show behavior similar to the critical point of a continuous transition. In practice spinodals are very hard to reach, because it implies to drive the system on the metastable solution, where any slight perturbation may take it to the stable one. Thus, we will now consider the case of 1D maps resembling hysteretic behavior.

\subsubsection{Extended subcritical pitchfork bifurcation}

The discrete form of the so-called extended normal form of the subcritical pitchfork bifurcation presents a jump  (discontinuous) bifurcation with hysteresis \cite[Chapter 11]{Chasnov}. For simplicity, we will consider the related map. Since the map is symmetric ($x\to -x$), only consider cases with $x\geq 0$.  The same results hold for $x<0$. The map equations are: 

\begin{equation}
    x_{l+1}=(1+\mu) x_l +x_l^3-x_l^5,
\end{equation}
and it has stable solutions $x_0=0$ for $\mu<0$ and $x_{\pm}=\pm \sqrt{\frac{1}{2}(1+\sqrt{1+4\mu})}$ for $\mu>-1/4$. Consequently, there are 2 stable solutions in the range $-1/4<\mu<0$ \cite{Chasnov}. Notice also that the solutions are discontinuous: $x_{\pm}(-1/4) \neq 0$, and also  $x_{\pm}(0) \neq 0$. The map also has a pair of unstable solutions in the range $-1/4<\mu<0$: $x_{u\pm}\pm \sqrt{\frac{1}{2}(1-\sqrt{1+4\mu})}$.

\begin{figure}[ht!]
    \centering
    \includegraphics[width=0.5\linewidth]{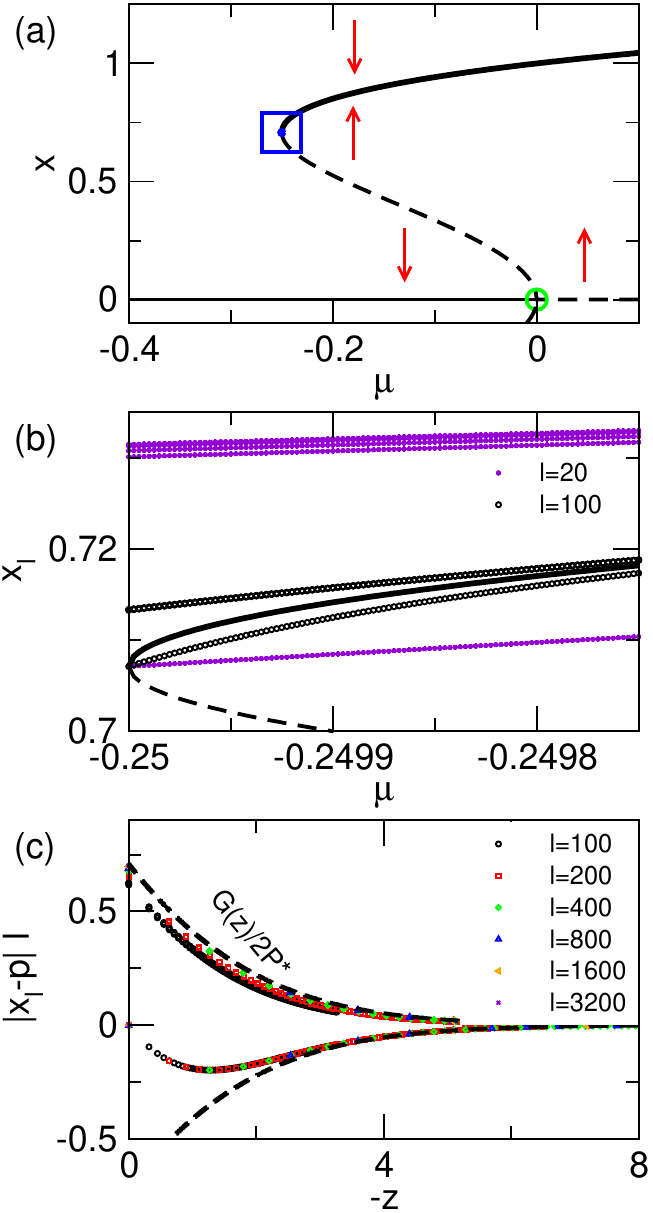}
    \caption{Stationary solutions and  collapse on a unique function after rescaling in the discrete form of the extended supercritical pitchfork map. Panel a: stationary solutions with $x\geq 0$ as a function of $\mu$. Dashed lines stand for unstable solutions. Red arrows indicate the flow of the solutions. green circle and blue square stand for the limit  points of lower and upper stable solutions.  Panel b: $x_l$ as a function of $\mu$ for different initial conditions about the upper limit point (blue square in panel a), after $l=20$ and $l=100$ steps. Notice that the range of $\mu$ and $x$ values is  smaller than in panel a. Panel c: function collapse for initial conditions about the upper limit point (blue square in panel a). In all panels, the initial conditions are: $x_0=0.8-0.95$ (collapse from above) and $x_0=1/\sqrt{2}$. }
    \label{fig:discont}
\end{figure}

In Fig. \ref{fig:discont}-(a), the stationary solutions are shown. Initial conditions starting above  the upper stable solution ($x=\sqrt{\frac{1}{2}(1+\sqrt{1+4\mu})}$), collapse well under the same functions used before (see panel c). When we attempt to collapse solutions starting below this curve, we face the problem that, as $\mu\to-1/4$ the basin of attraction of the upper stable solution gets narrower, collapsing to a single point at $\mu=-1/4$, so, the only possible initial condition is $x_0=1/\sqrt{2}$, which is also the final condition at $\mu=-1/4$  (see panel b). As a consequence, similar to the spinodals in discontinuous transitions, a collapse is theoretically present but practically unobservable. A similar limitation holds for the lower stable solution, $x=0$, as we reach the other ``spinodal" $\mu\to 0^-$.

\subsection{2D maps}

We now turn to 2D maps to examine how finite-time scaling generalizes beyond one-dimensional dynamics. In simple cases, a change of variables can decouple the system so that the bifurcation involves only one transformed variable, effectively reducing the dynamics to a 1D map where previous results apply. For more general 2D systems, however, analytical treatment becomes intractable. In one dimension, the scaling variable $z = l(m - 1)$ depends on the local slope $m = f'(p)$, which characterizes convergence near a fixed point. In higher dimensions, a natural generalization is to use the Jacobian matrix $J$ evaluated at the fixed point. Since $\det J$ measures local volume contraction or expansion, we propose that $m$ be replaced by $\sqrt[n]{\det J}$, where $n$ is the system’s dimensionality. This definition preserves the multiplicative structure of linear flows and reduces to $m$ in the 1D limit. In two dimensions, this becomes $\sqrt{\sigma_1 \sigma_2}$, where $\sigma_1$ and $\sigma_2$ are the Jacobian eigenvalues. We test this hypothesis numerically on 2D maps of increasing complexity and find that finite-time scaling collapse remains qualitatively consistent with this choice. Finally, we apply this reasoning to the full 2D Chialvo neuron map, where strong coupling between variables prevents reduction to an effectively one-dimensional form.

\subsubsection{Supercritical Hopf bifurcation}

\begin{figure}[ht!]
    \centering
    \includegraphics[width=0.95\linewidth]{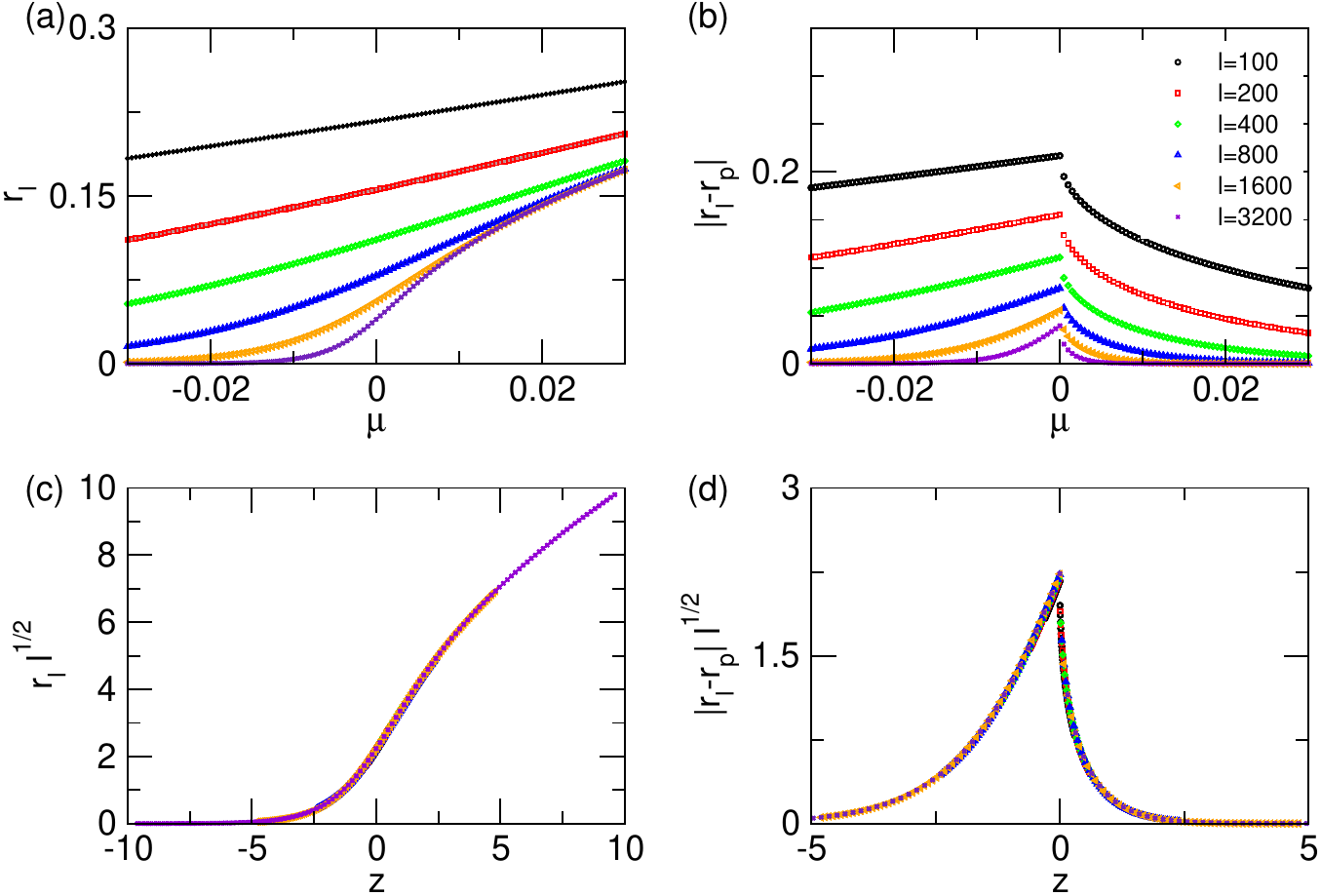}
    \caption{Stationary solutions and collapse on a unique function after rescaling in Hopf bifurcation. Panel a: solutions $r_l$ for different values of $l$, as a function of $\mu$. Panel b: Distance to the closest fixed point. Panel c: collapse after rescaling of the distance to the midpoint $p_m$, plotted as a function of $z$. Panel d: Collapse after rescaling of the distance to the closest function. }
    \label{fig:Hopf}
\end{figure}

We first consider a discrete form of the supercritical Hopf bifurcation, where a stable fixed point loses stability and gives rise to a stable limit cycle—a hallmark of oscillatory behavior in physical systems. Following the rationale in \cite{corral2}, near the bifurcation, we write a differential equation  as a map of discrete, infinitesimally small, increments of time, $\delta t$. We thereby obtain to the following discretized version of the Hopf map: 

  \begin{eqnarray} 
  f_{\mu}^{(x)}=x+dt(\mu-x^2-y^2)x-{dt}( \omega y)\\
  f_{\mu}^{(y)}=y+dt(\mu-x^2-y^2)y+{dt}(\omega x) \label{Hopf}
  \end{eqnarray}
where $dt\ll 1$, and $\omega$ is an angular velocity. We can make a change of variables towards $r=\sqrt{x^2+y^2}$ and an angular variable $\theta$.  The evolution equation for $r$ is now independent of $\theta$, and destabilizes as an 1D map: it is given by $f_{\mu}^{(r)}=r+dt(\mu-r^2)r$, while the role of $\theta $ becomes irrelevant. The  approach  for 1D maps can be applied to $r$,   see Fig. \ref{fig:Hopf}. A collapse in the 3 dimensional space $\mu l$, $\sqrt{l}x$, $\sqrt{l}y$ is shown in Fig.\ref{Hopf3D}.

\subsubsection{2D Chialvo neuron Map }
Having validated the generalized scaling structure in simplified 2D examples, we now turn to the full 2D Chialvo neuron map as a biologically inspired, nontrivial test case.
The model is inspired in the coupled map lattice numerical approach \cite{kaneko} which considers time and space as discrete variables but state as a continuous one.   The map is able to efficiently mimic generic neuronal dynamics in computational simulations, as single elements or as parts of inter-connected networks. The evolution equations  are:
\begin{eqnarray}
    x_{l+1} &=& x_l^2 e^{y_l-x_l}+k \nonumber \\
    y_{l+1}&=&a y_l -b x_l +c,\label{eq:Chialvo2D}
\end{eqnarray}
where $x$ acts as the so-called activation (or potential or voltage-like) variable and $y$ as a recovery-like variable. Subscripts $l$ represent iteration steps corresponding to the discretized time evolution of the system.  The model includes four parameters. In the activation variable, the parameter $k$ can act either as a constant bias or as a time dependent additive perturbation. For $x=0$, the fixed point of the recovery variable ($y_f$) is determined by three positive parameters: $a$, the time constant of recovery ($a < 1$); $b$, the activation-dependence of the recovery process ($b < 1$), and the offset  $c$.
Typical parameters are $a=0.89$, $b=0.6$ $c=0.28$. Under such conditions, the model exhibits a non-autonomous regime with normal excitability, where an external perturbation generates spikes for low values of $k$ ($k\sim 0.02$). 
For relatively larger values of $k$ ($k\sim 0.029$) a non-autonomous regime with supernormal excitability and autonomous (i.e. always spiking solutions) for even larger $k$ values ($k\sim 0.03$). At this point, the dynamics is bi-stable, having both oscillatory and fixed point solutions (see \cite{Chialvomap} for a more detailed description).
 
\begin{figure}
    \centering
    \includegraphics[width=0.95\linewidth]{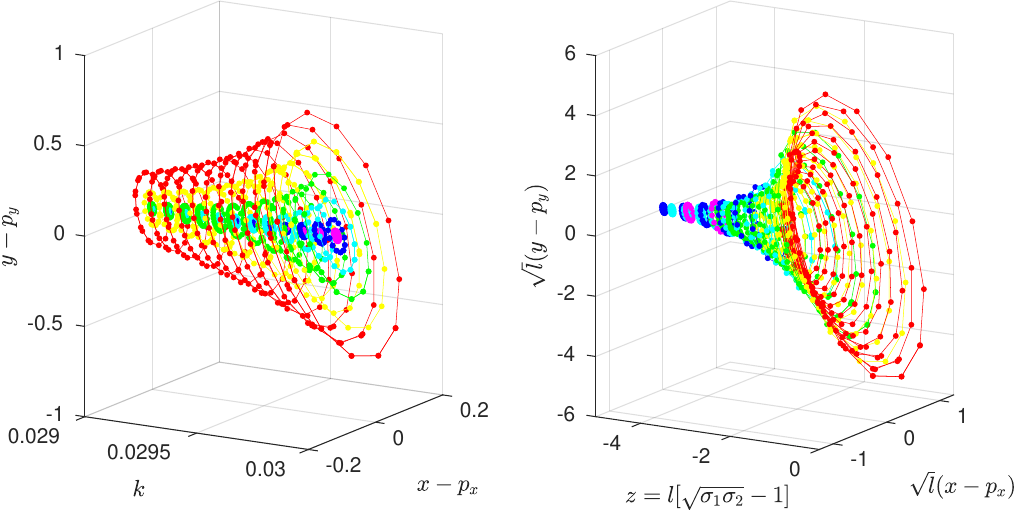}
    \caption{Approximate collapse  for 2D Chialvo map. Left: $x$ and $y$ as a function of $k$ for $l=50$ (red),  $l=50\sqrt{2}$, $l=100$,..., $l=282$. Right: 
    $\sqrt{l} (x-p_x)$ and $\sqrt{l} (y-p_y)$  as a function of $l (\sqrt{\sigma_1\sigma_2}-1)$, being $\sigma_i$ the eigenvalues at the point $[p_x,p_y]$. Results for $a=0.89$, $b=0.6$ $c=0.28$ and different values of $k$ in the range $k=0.029$ to $k=0.0299$. Only properly chosen initial conditions, converging to the no spiking solution of the non-autonomous regime were considered.  In both panels, the trajectory of each point is normalized to share the same initial distance to the fixed point.}
    \label{fig:ch2d}
\end{figure}

 Influenced by the results for 1D discontinuous maps, we will  attempt to collapse the solutions that converge to the lower ``branch" (the non spiking one). Results are shown in Fig. \ref{fig:ch2d}. For this case, initial conditions have to be chosen carefully, assuring that they are inside the separatrix enclosing the non-spiking solutions. For each value of $\mu$, one initial condition was fixed, and the following initial conditions were generated  evolving the previous initial condition 5 steps, and dividing the resulting point by the square root of the product of the eigenvalues to the fourth power, and so on. Also, to yield reasonable comparisons among different $\mu$ values, we had to normalize each family of trajectories by the average distance to the  fixed point.

\section{Discussion}\label{conclusion}

We extended the study of finite-time scaling to several classes of bifurcations in both one- and two-dimensional maps. Starting from period-doubling bifurcations in the logistic map and the reduced 1D Chialvo neuron model, we derived explicit scaling laws with critical exponents determined by the local nonlinearity $k$. For discontinuous (first-order-like) bifurcations, we adapted the framework to spinodal-type transitions, generalizing the results of Refs.~\cite{corral1,corral2} and providing closed-form expressions for the relevant coefficients and scaling variables. We then examined two-dimensional maps exhibiting bifurcations from fixed points to periodic orbits. While full analytical treatment is generally intractable in two dimensions, we found that finite-time scaling remains robust when the linear term $m$ is replaced by the square root of the dominant Jacobian eigenvalue—suggesting that the scaling structure may extend across dimensions and bifurcation types. These results naturally lead to questions about higher-order bifurcations, where the number of solutions increases and chaotic behavior may emerge. Similar extensions could be explored in more complex 2D maps and in continuous-time systems.

Beyond its technical applications, finite-time scaling offers a deeper physical interpretation. Although developed for low-dimensional systems, its structure mirrors key features of critical phenomena in statistical physics: (1) scaling exponents are nontrivial and governed by $k$, akin to universality classes; (2) time serves as an effective system size, with the scaling variable $z = l(m - 1)$ analogous to $L^{1/\nu}(T - T_c)$ in finite-size scaling; and (3) the framework extends beyond trajectory collapse to parametric sensitivity, with $\chi_l(\mu)$ behaving like a dynamical susceptibility closely related to Fisher information under Gaussian assumptions. In passing, please note that the finite time susceptibility mirrors the behaviour of the finite \emph {size} susceptibility demonstrated in the analysis of midges swarms (e.g. Fig. 2a of Ref.\cite{attanassi}) as well as for the dynamics of proteins (see Fig. 3a of Ref.\cite{tang}).

A hallmark of criticality in physical systems is the simultaneous emergence of multiple power-law behaviors linked by scaling relations, reflecting deep coherence among fluctuations, responses, and relaxation times. Our results reveal similar structure: both $x_l - p$ and $\chi_l(\mu)$ follow finite-time scaling laws governed by the same $k$, with exponents satisfying predictable relations. This suggests that even simple low-dimensional systems can briefly emulate the collective dynamics of more complex high-dimensional systems. In this sense, bifurcation analysis serves as a dynamical analogue of critical-point theory, revealing effective order parameters whose transient evolution captures deeper statistical structure. 

 While continuous transitions in critical phenomena often involve an ordered and a disordered phase, the examples considered here undergo  transitions among two stable solutions,  resembling more a transition between two ordered phases. A possible direction for future work is to extend finite-time scaling to  maps at the edge of chaos.

Finally, while low-dimensional maps offer analytical and computational tractability, they inevitably oversimplify the rich behavior of more complex systems—for instance, by averaging over spatial correlations or ignoring high-dimensional interactions.    A key direction for future work is to understand how phase transitions in high-dimensional systems—whether continuous or discontinuous—manifest as finite-time scaling in reduced models. In particular, identifying whether response observables retain signatures of full-system dynamics may help bridge the gap between model reduction and statistical complexity. Notably, in many real-world contexts—ranging from ecosystem collapse to neuronal activity—transitions are preceded by finite-time trends such as slowing down, increased sensitivity, or heightened fluctuations, which are known to be linked  through dynamic scaling relations (see e.g. \cite{reviewHH77,reviewGrigera,KPZ,swarms,Camargo}).
The observables studied here, including finite-time susceptibility and Lyapunov exponents, naturally capture such precursors and may serve as quantitative indicators of proximity to bifurcation. This connection suggests that finite-time scaling could provide a unifying perspective on early-warning diagnostics across disciplines, highlighting how critical-like scaling emerges in simplified settings and motivating further efforts to link reduced dynamics with collective transitions in complex systems.

\section*{Acknowledgments}
QYT thanks Natural Science Foundation of China (No. 12305052), Research Grants Council of Hong Kong (Nos. 22302723 and SRFS2324-2S05), and Hong Kong Baptist University's funding support (RC-FNRA-IG/22-23/SCI/03).
DRC thanks Hong Kong Baptist University for funding his Distinguished Professorship of Science during these studies. 
\vspace{1cm}
\hrule 

\appendix

\setcounter{figure}{0}  
\renewcommand\thefigure{A. \arabic{figure}}

\section{Derivations}

In this appendix, we derive the finite-time scaling behavior underlying the main results. We first analyze simple one-dimensional maps with period-doubling bifurcations, showing how symmetry determines the effective scaling exponent. We then examine the scaling of susceptibility and Lyapunov exponents, and extend the framework to the logistic and reduced Chialvo maps. Finally, we illustrate collapse behavior in a two-dimensional Hopf bifurcation.

\subsection{Simple map  }

In this appendix, we provide analytical justification for the finite-time scaling behavior observed in a class of simple one-dimensional maps introduced in Eq.~\ref{mapk}, which serve as minimal models for studying period-doubling bifurcations near $\mu = 0$. We examine how the nonlinearity exponent $\kappa$ in the original map gives rise to an effective scaling exponent $k$, with the relationship depending on the map’s symmetry. These results form the theoretical basis for the scaling collapses presented in the main text.
\begin{figure}[ht!]
    \centering
    \includegraphics[width=0.95\linewidth]{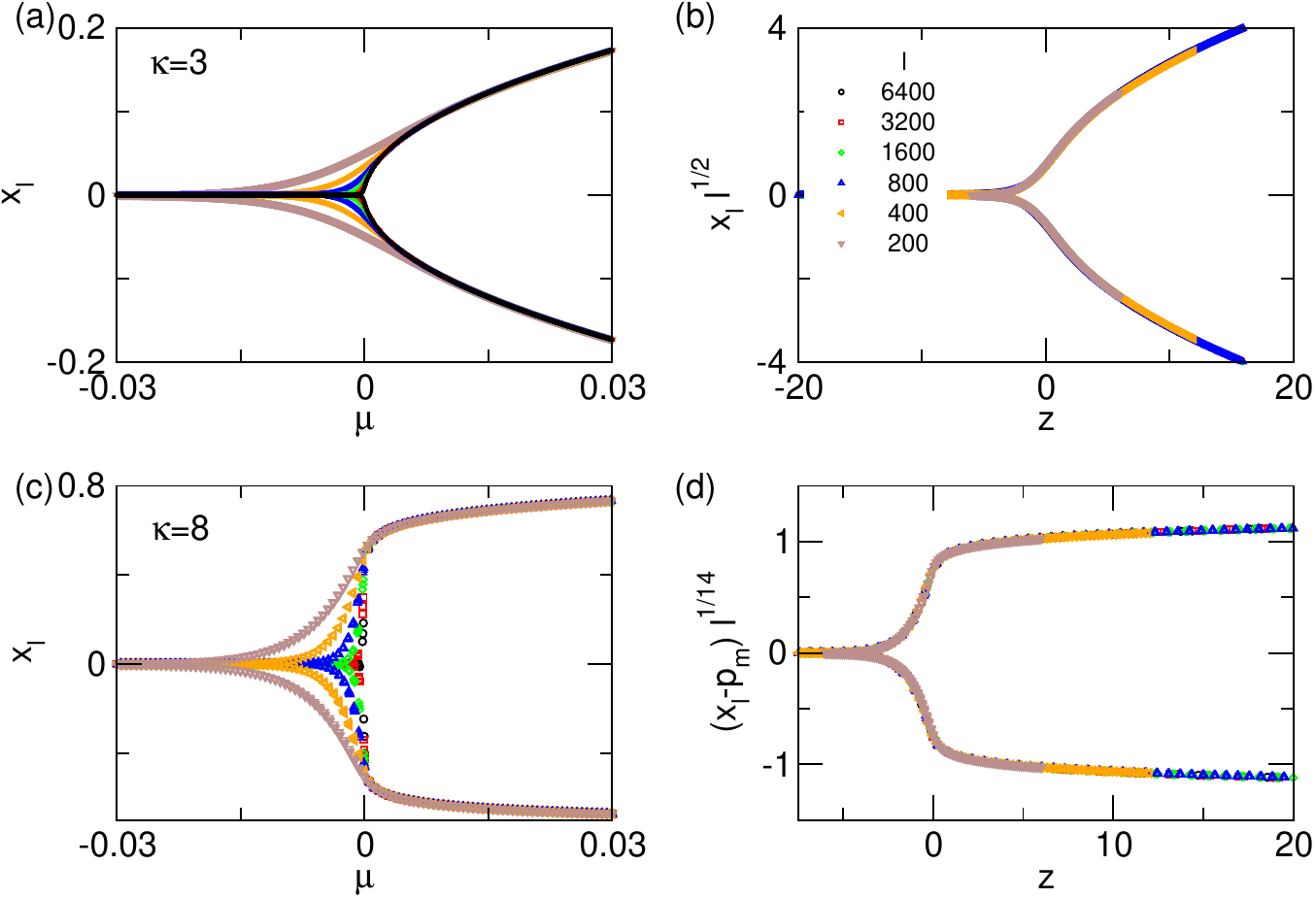}
    \caption{Results for simple period-doubling maps,  {$f_{\mu}(x)=-(1+\mu)x+ x^\kappa$, for $\mu\simeq 0$}.  Panel a: solutions for $\kappa=3$ as a function of $\mu$. Panel b: data collapse. Panels c and d: same as (a) and (b), for $\kappa=8$. Here $z=\mu l$. }
    \label{fig:Simple}
\end{figure}

\textbf{For  $\kappa$ odd:}
Let us consider $y_l=|x_l|$, from Eq. \ref{mapk}, we define $g$ such that $y_{l+1}=g(y_l)$. It is straightforward to find that   $$ g(y)=m y-Q y^\kappa,$$ 
That is, $g$ is a map already considered by Eq. \ref{general}, but now $g'(0)=m=1+\mu=-f'(0)$. The distance to the \emph{stable solution} is given by already derived results (Eq. \ref{Eqk}). The same holds for $\mu>0$, when we consider the distance to the unstable solution, $y=0$, which, in the general case, is the distance to the average of the period-2 solutions.

To collapse the distance to the \emph{closest solution}, we may write $(x_l-p) l^{\frac{1}{k-1}}=[x_l-p_m] l^{\frac{1}{k-1}} -[p_m-p] l^{\frac{1}{k-1}}$. The first term between brackets is essentially what we have been plotting, and the second term  is $\sqrt[k-1]{\mu/Q}$, which after multiplying by $l^{\frac{1}{k-1}}$ is equal to $z^{\frac{1}{k-1}}$.

\textbf{For  $\kappa$ even:}
The relation $f(-x)=-f(x)$ no longer holds, so we need to consider $f(f(x))$. It is straightforward to find that
\begin{eqnarray}
    x_{l+2}&=&m^2x_l+Q(m+m^\kappa)x_l^\kappa+(\kappa-1)Q^2 m^{\kappa-1}x_l^{2\kappa-1}\nonumber \\&+&O(x_l^{3k-2}), \label{Par}
\end{eqnarray}
where now $m=-1-\mu$. For $\mu\to 0$, The linear term is $m^2\simeq 1+2\mu$. Since $\kappa$ is even, the next term $m+m^\kappa\simeq (\kappa-1)\mu \to 0$. The following term is  nonzero. The coefficient tends to $-(\kappa-1)Q^2$, and the exponent is $2\kappa-1$. Then $f^2$ and consequently $f$, are well described by Eq.\ref{general}, taking $k=2\kappa-1$ {and $P^*=Q^2 (\kappa -1)$}, replacing the distance to the stable solution by the distance to the midpoint $p_m$, for $\mu>0$.

In Fig. \ref{fig:Simple}, we illustrate this results for one odd vale of $\kappa$ ($\kappa=3$) and one even value ($\kappa=8$).

\subsection{Applicability to Logistic and Chialvo Maps}

For odd $\kappa$, the map directly follows the form of Eq.~\ref{general}, while for even $\kappa$, the dominant term emerges from the composition $f(f(x))$, leading to an effective exponent $k = 2\kappa - 1$. In both cases, the susceptibility scales as
\[
\chi_l \sim l^{(3k - 5)/(k - 1)},
\]
with $k = \kappa$ (odd) or $k = 2\kappa - 1$ (even). We now show how this result applies to two representative models used in the main text.

\noindent\textbf{Logistic Map.}  
The logistic map $f_{\mu}(x) = \mu x(1 - x)$ does not match the canonical forms above. However, expanding around the bifurcation point $\mu^* = 3$ and the fixed point $x^* = 2/3$, we obtain:
\[
f_{\mu}(x) \approx x - (\mu - 3)(x - 2/3) - \mu(x - 2/3)^2 + \dots
\]
This yields an effective cubic nonlinearity with $k = 3$. Inserting into the general susceptibility formula gives:
\[
\chi_l \sim l^{(3 \cdot 3 - 5)/(3 - 1)} = l^{4/2} = l^2.
\]
The observed exponent in simulations consistent with this prediction.  

\vspace{0.5em}
\noindent\textbf{Reduced 1D Chialvo Map.}  
For the reduced Chialvo map $f_{\mu}(x) = x^2 e^{\mu - x}$, we perform a Taylor expansion near the critical point $x^* = 3-\log(3)$:
\[
f_{\mu}(x) \approx f_{\mu}(x^*) + f'_{\mu}(x^*)(x - x^*) + \frac{1}{2} f''_{\mu}(x^*)(x - x^*)^2 + \dots
\]
The linear coefficient satisfies $f'_{\mu}(3) = -3 e^{\mu - 3}$, and the dominant nonlinearity is quadratic, and period-doubling, so $\kappa = 2$,then $k=3$. Substituting into the general form:
\[
\chi_l \sim l^{(9 - 5)/(3 - 1)} = l^2.
\]

\begin{figure}
    \centering
\includegraphics[width=0.95\linewidth]{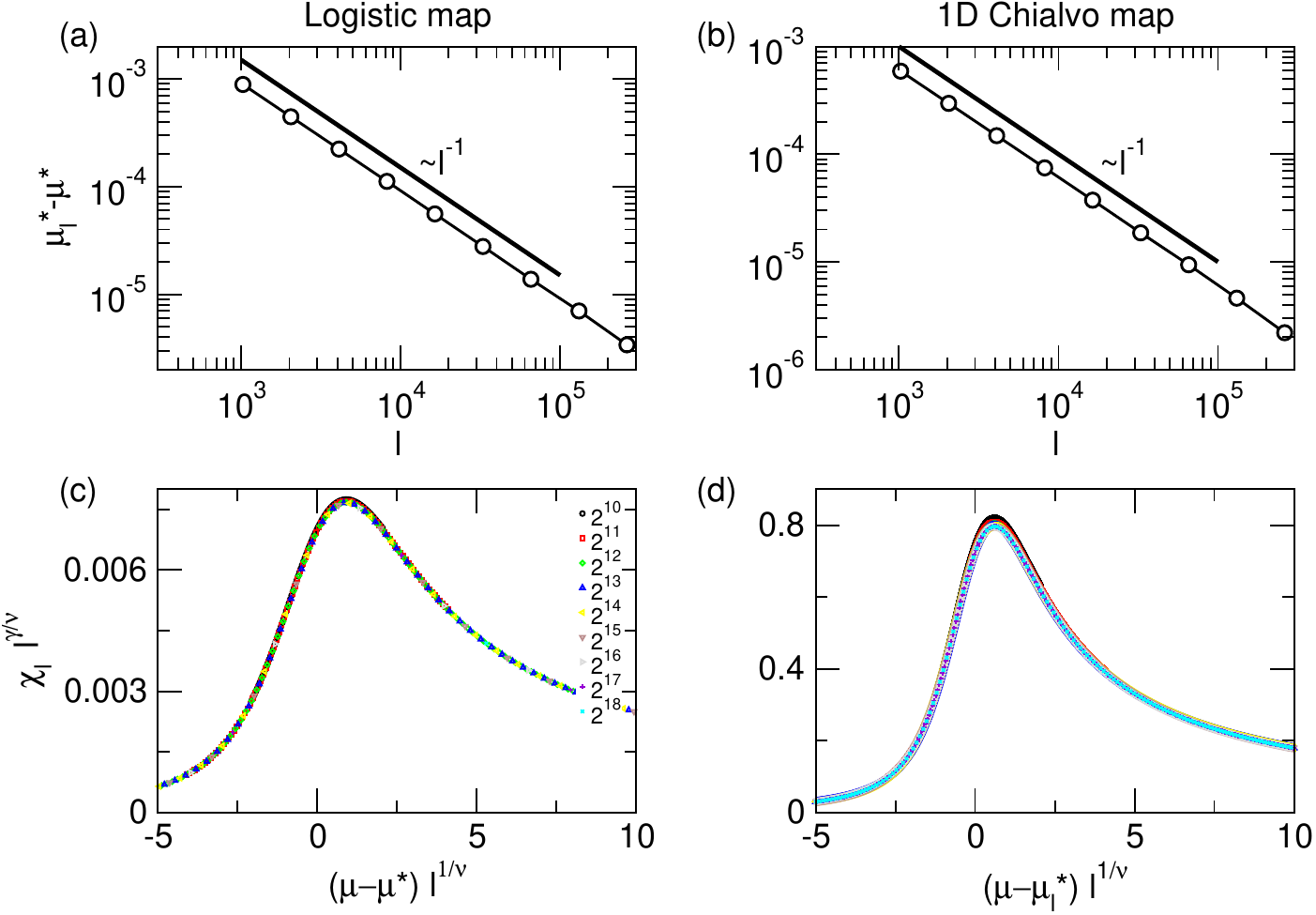}
    \caption{Scaling and Collapse of the finite time susceptibility for the Logistic and the 1D Chialvo maps. Panel (a): $\mu_l^*-\mu^*$ as a function of $l$ for the Logistic map. Panel b: same results for the 1D Chialvo map. Panel c: Susceptibility collapse for the Logistic map, considering $\gamma=2$, $\nu=1$. Panel d: same as panel c for 1D Chialvo map.
    All parameters as in Fig. \ref{fig:Exp1}.}
    \label{Colapse}
\end{figure}
We now consider the distance of among the $\mu_l^*$, the value of $\mu$ that maximizes $\chi_l$ and $\mu^*=\lim_{l\to\infty} \mu_l^*$. We find numerically that $|\mu_l^*-\mu^*|\propto l^{-1/\nu}$ with $\nu=1$. The scaling of the peak of $\chi_l$ together with the scaling of $\mu_l^*$ suggests that a collapse of the curves may take place, in analogy to statistical thermodynamics results. We find that this is the case indeed, see Fig. \ref{Colapse}
\vspace{0.5em}
\begin{figure}[ht!]
    \centering
    \includegraphics[width=0.95\linewidth]{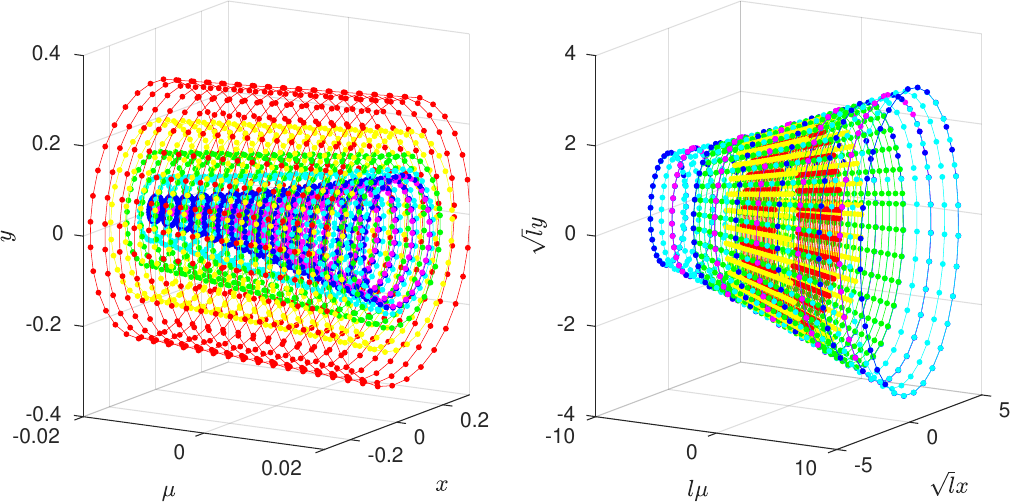}
    \caption{3D view of Hopf Bifurcation collapse for $dt=0.1$ and $\omega=0.01$. Left: $x$ and $y$ as a function of $\mu$ for $l=50$ (red),  $l=100$,..., $l=1600$. Right: Curve collapse. Initial conditions consist of points on the unit circle ($x_0^2+y_0^2=1$).}
    \label{Hopf3D}
\end{figure}

\subsection{3D plot of Hopf Collapse}

Although the logistic and Chialvo maps are not of the exact canonical form, they can be locally approximated by polynomial maps with identifiable effective nonlinearity $k$. The general susceptibility scaling $\chi_l \sim l^{(3k - 5)/(k - 1)}$ remains valid as long as the local expansion is dominated by a single nonlinear term.


\newpage
\begin{thebibliography}{99}

\bibitem{may} R. May. Simple mathematical models with very complicated dynamics.\emph{Nature} 261, 459-467 (1976) 

\bibitem{ott}E. Ott. Chaos in dynamical systems. Cambridge Univ. Press (1993) 

 \bibitem{kaneko} K. Kaneko. Theory and applications of coupled map lattices. Nonlinear science: theory and applications. John Wiley \& Sons (1993)

 \bibitem{kaneko2} Kaneko, K. Period-Doubling of Kink-Antikink Patterns, Quasiperiodicity in Antiferro-Like Structures and Spatial Intermittency in Coupled Logistic Lattice: Towards a Prelude of a ``Field Theory of Chaos''.\emph{Progress of Theoretical Physics}, 72 (3) 480-486 (1983)
 
\bibitem{glass}L. Glass \& M.C. Mackey. From Clocks to Chaos: The Rhythms of Life. Princeton Univ. Press, (1988)


\bibitem{review1}B. Ibarz, J.M. Casado, M.A.F. Sanjuán, Map-based models in neuronal dynamics, \emph{Physics Reports}, 501 (1–2): 1-74 (2011)

\bibitem{review2}
M. Girardi-Schappo, M.H.R. Tragtenberg, O. Kinouchi,
A brief history of excitable map-based neurons and neural networks,
\emph{J. of Neuroscience Methods}, 220 (2):116-130 (2013)

\bibitem{corral1} A. Corral, L. Alsedà,  J. Sardanyés. Finite-time scaling in local bifurcations. \emph{Sci. Rep.}, 8(11783), (2018)

\bibitem{corral2} A. Corral. Universal finite-time scaling in the transcritical, saddle-node, and pitchfork discrete and continuous bifurcations. \emph{Chaos: An Interdisciplinary Journal of Nonlinear Science} 35 (1) (2025)

\bibitem{falco}C. Falcó, A. Corral. Finite-time scaling for epidemic processes with power-law superspreading events. \emph{Phys. Rev. E} 105 (6), 064122 (2022)

\bibitem{WG1} R. Garcia-Millan, F. Font-Clos,  A. Corral. Finite-size scaling of survival probability in branching processes. \emph{Phys. Rev. E}, 91:042122, (2015)

\bibitem{WG2} A. Corral, R. Garcia-Millan,  F. Font-Clos. Exact derivation of a finite-size scaling law and corrections to scaling in the geometric Galton-Watson process.\emph{PLoS ONE}, 11(9):e0161586, (2016)

\bibitem{Chialvomap} D. R. Chialvo. Generic excitable dynamics on a two-dimensional map.  \emph{Chaos, Solitons \& Fractal}, 5 (3): 461–479, (1995)

\bibitem{footnote} For $Q>0$ and/or $\kappa$ even, there is also an unstable solution $\sqrt[k-1]{2+\mu \over Q}$.

\bibitem{Strogatz} S. H. Strogatz. Nonlinear dynamics and chaos. 2nd ed. CRC Press (2018)

\bibitem{Chasnov} J. R. Chasnov. Applied Linear Algebra and Differential
Equations. LibreTexts (2025)

\bibitem{footnote2}Recall that the fixed point satisfies $p e^{\mu - p} = 1$, and the bifurcation occurs when $f'(p^*) = -1$.

\bibitem{attanassi} A. Attanasi, A. Cavagna, L. Del Castello, I. Giardina, S. Melillo, L. Parisi, O. Pohl, B. Rossaro, E. Shen, E. Silvestri, M. Viale, Finite-Size Scaling as a Way to Probe Near-Criticality in Natural Swarms, \emph{Phys. Rev. Lett.} 113, 238102 (2014)

\bibitem{tang}Q.Y. Tang, Y.Y. Zhang, J. Wang, W. Wang, D.R. Chialvo. Critical Fluctuations in the Native State of Proteins,
\emph{Phys. Rev. Lett.} 118, 088102 (2017)

\bibitem{reviewHH77}P. C. Hohenberg, B. I. Halperin, Theory of dynamic critical phenomena, \emph{Rev. Mod. Phys.} 49, 435 (1977)

\bibitem{reviewGrigera} T.S. Grigera, Correlation functions as a tool to study collective behaviour phenomena in biological systems, \emph{J. Phys. Complex.} 2, 045016 (2021)

\bibitem{KPZ}   M. Kardar, G. Parisi, Y.-C. Zhang, Dynamic Scaling of Growing Interfaces, \emph{Phys. Rev. Lett.} 56, 889 (1986)


\bibitem{swarms}A. Cavagna, D. Conti, C. Creato, L. Del Castello, I. Giardina, T.S. Grigera, S. Melillo, L. Parisi, M. Viale, Dynamic scaling in natural swarms. \emph{Nature Phys.} 13, 914–918 (2017)

 
\bibitem{Camargo} S. Camargo, D. A. Martin, E. J. A. Trejo, A. de Florian, M. A. Nowak, S. A. Cannas, T. S Grigera, D. R. Chialvo, Scale-free correlations in the dynamics of a small  ($N=10000$) cortical network, \emph{Phys. Rev. E} 108, 034302
(2023)


\end{thebibliography}
\end{document}